\documentstyle[12pt,oldlfont]{article}
\font\bbf=cmbx10 scaled\magstep1
\newfont{\bldm}{cmmib10}
\def\sigbol{\mbox{\bldm\symbol{"1B}}}
\def\taubol{\mbox{\bldm\symbol{"1C}}}
\def\phibol{\mbox{\bldm\symbol{"1E}}}
\def\half{{\textstyle{1\over 2}}}
\def\shalf{{\scriptstyle{1\over 2}}}
\def\rvec{{\bf r}}
\def\kvec{{\bf k}}
\def\altext#1#2{{\textstyle{#1\over #2}}}
\def\K{{\scriptscriptstyle K}}
\def\N{{\scriptscriptstyle N}}

\begin{document}
\begin{flushright} MC/TH 95/16\\ hep-ph/9509366 \end{flushright}

\vskip 5pt
\centerline{\bbf Magnetic Moments of the Octet Baryons}
\centerline{\bbf in the Colour-Dielectric Model}
\vskip 30pt
\centerline{Moo-Sung Bae
and Judith A. McGovern\footnote{Electronic address j.mcgovern@man.ac.uk} }
\vskip 20pt
\centerline{\it Theoretical Physics Group}
\centerline{\it Department of Physics and Astronomy}
\centerline{\it University of Manchester}
\centerline{\it Manchester, M13 9PL, U.K.}
\vskip 50pt
\centerline{\bbf Abstract}
\vskip 12pt
Baryon magnetic moments are calculated in the colour-dielectric model with
pion and kaon loops. The only free parameter of the model is determined from
the nucleon isoscalar radius, and all SU(3) symmetry breaking, including that
in the quark sector, is determined by mesonic masses and decay constants.
Good agreement with experiment is obtained for the ratios of the magnetic
moments, but the inclusion of kaons does not improve the results.
The results obtained in this approach are significantly better than any that
have been obtained in hedgehog-based models.

\newpage

\section{Introduction}
A satisfactory description of the magnetic moments of baryons
within the framework of a dynamical model has proved harder to obtain
than might have been expected.  Although a naive additive quark model with
different strange and non-strange magnetic moments reproduces the overall
pattern reasonably well, there are some notable problems, such as fitting
$\mu_p-\mu_{\Sigma^+}$ at the same time as $\mu_\Lambda$ and $\mu_\Omega$,
or obtaining the correct sign of $\mu_{\Xi^-}-\mu_\Lambda$. However very
few dynamical models come close to
doing even as well. The problem of the magnetic moments is
interesting precisely because it is appears to be a sensitive test of the
inclusion of SU(3)-breaking effects beyond first order.
Numerous attempts in Skyrme and other hedgehog-based
models have all failed to improve on the
additive quark model.\ \cite{Chemtob, Nyman, Oh, Park, Park1}

One of the best results within the framework of a
relativistic quark model was obtained twelve years ago by Th\'eberge and
Thomas\ \cite{T&T} in the cloudy bag model with pion loops included
perturbatively.  A more recent model which, like the cloudy bag model, has
a relatively weak quark-pion coupling, but which unlike it is fully
dynamical, is the colour-dielectric model\ \cite{CD}.
  In this model the role of the
static bag is taken by a scalar, chiral singlet field which confines the
quarks.  Chiral invariance is restored by a implementing the coupling
through the $\sigma$ field of the linear $\sigma$ model, with a
matching coupling to pions.  Though  hedgehog solutions can be found,
the $\sigma$ and pion fields remain near their vacuum values\
\cite{chiral CD}.  Thus a cloudy bag approach to the inclusion of mesons is
more natural\ \cite{W&D}.  This approach, including pion and kaon fields,
has been successfully used to calculate the octet and decuplet mass
spectra, in an approach in which the strength of the SU(3) symmetry breaking
was governed entirely by the physical pion and kaon masses and coupling
constants\ \cite{JMCD}. The current paper applies the same method to
the calculation of the magnetic moments.

The paper is organised as follows.  In section~2 the colour-dielectric model is
introduced and the perturbative inclusion of mesons explained.  In section~3
the various contributions to the magnetic moments are derived. In these two
sections excessive overlap with refs.~\cite{T&T,JMCD} has been avoided  where
possible. In section~4 the results are presented and a comparison with other
models and with the results of chiral perturbation theory is given.

\section{The colour-dielectric model}

The version of the colour-dielectric model used in this paper is the
three-flavour one developed by McGovern \cite{JMCD}.
The original model involves quarks and a $\chi$ field representing a
glueball or glueball-meson hybrid, with an unusual interaction term:
\begin{equation}
{\cal L}={\overline\psi}\left[i\gamma^\mu D_\mu
+{g\over\chi}\right]\psi +\half\partial_\mu \chi \partial^\mu \chi -U(\chi)
+\kappa(\chi)F^{\mu\nu}F_{\mu\nu}.
\end{equation}

The potential $U(\chi)$ has a global minimum at $\chi=0$, so $\chi$ vanishes
in the vacuum.
The covariant derivative and the field tensor $F^{\mu\nu}$ involve
``coarse-grained" gluons, which are to be treated as Abelian and only included
to lowest order, since non-perturbative effects are included through the
$\chi$ field.  The colour dielectric, $\kappa$, is proportional to $\chi^4$
and thus vanishes in the vacuum, ensuring that colour is confined.  The inverse
quark-$\chi$ coupling provides confinement even for
colourless quarks, as their
effective mass in the vacuum would be infinite.

Many groups have now obtained soliton solutions in this model, either ignoring
the gluons or including one-gluon exchange self-consistently or perturbatively.
It has been shown\ \cite{JMCD} that, at least in the case of a simple quadratic
$\chi$ potential, first-order perturbation theory works well.

This model is of course not chirally symmetric.  It can be made so by allowing
the quarks to couple to the inverse $\chi$ field only through an appropriate
combination of meson fields---in the SU(2) case, $g/\chi\to g(\sigma+i\gamma_5
\taubol\cdot\phibol)/(f_\pi \chi)$ where $\phibol$ is the pion isotriplet and
$\sigma$ its scalar, isoscalar chiral partner; in this linear sigma model
the hidden nature of chiral symmetry is reflected in a vacuum expectation value
of $f_\pi$ for the $\sigma$ field.  When these mesons are included in a
hedgehog ansatz the deviation from their vacuum expectation values within the
soliton is small; it is therefore possible to ignore them in the construction
of
the soliton and include pion contributions to observables perturbatively,
in the manner of the cloudy bag model.

Three flavours and broken SU(3) symmetry can also be incorporated in the model.
(For details, see ref.~\cite{JMCD}.)
Formally this requires two nonets, of scalars and pseudoscalars; the meson
potential is a generalised mexican hat with explicit symmetry-breaking terms
which generate separate pion and kaon masses, and two of the scalar fields, the
singlet and the eighth member of the octet, develop vacuum expectation values
which are determined by the pion and kaon decay constants.  If the same
strategy
of ignoring deviations of the meson fields from their vacuum expectation values
is adopted, one is left with a very simple model in which strange and
non-strange quarks couple to the inverse $\chi$ field with different coupling
constants, the ratio of which is fixed:
\begin{equation}
c_g\equiv{g_s\over g_n}={2f_\K\over f_\pi}-1.
\label{cg}
\end{equation}
Light mesons, that is pions, kaons and the $\eta$, can be included
perturbatively. The interest of this model is that SU(3) symmetry breaking
effects in the  baryon sector are not  dependent on some arbitrary
strange-quark mass, but on well determined  experimental quantities from the
meson sector, $f_\K/f_\pi$ and $m_\K$.

The basic quark-$\chi$ Lagrangian is as follows:
\begin{equation}
{\cal L}={\overline\psi}\left(i\gamma^\mu \partial_\mu
+{{\bf G}\over\chi^p}\right)\psi +\half\partial_\mu \chi \partial^\mu \chi
-\half M_\chi^2\chi^2,
\label{lagrange}\end{equation}
where ${\bf G}$ is a matrix in flavour space, ${\bf G}={\rm
diag}(g_n,g_n,g_s)$,
and the possibility of varying the power $p$ of the inverse $\chi$-field which
couples to the quarks has been included. (Theoretical arguments can
be advanced for several values of the power $p$, but we will only
consider $p=1$ or 2.)   This Lagrangian has two free
parameters, $M_\chi$ and the dimensionless parameter
$\beta^2=g_n M_\chi^{-(p+1)}$
The standard procedure is to find the
solution for three non-strange quarks in the same s-wave orbital, and to fit
$M_\chi$ by
requiring that the r.m.s quark radius equal the nucleon isoscalar radius,
since that will be least affected by mesonic corrections. It is then
found that the energy and other properties are remarkably insensitive to the
value of $\beta$, and that $M_\chi$ scales as $\beta^{2/2p+1}$ at least for
$M_\chi$ greater than around 1~GeV.  This scaling has been shown to be exact
if the $\chi$ kinetic energy can be ignored \cite{JMscale}.

The average of the nucleon and $\Delta$ energy, as given by the flavour
symmetric calculation, is extremely high:
$E=1906$~MeV for $p=1$ and $E=1773$~MeV for $p=2$. However estimates of the
mesonic, gluonic and centre of mass corrections
bring these energies down considerably.

Baryons with non-zero strangeness are based on bare solitons with one or more
of the three quarks in an s-wave strange quark orbital.  The full octet and
decuplet mass spectrum has been calculated in ref.~\cite{JMCD}, and good
agreement with experiment obtained.  This paper explores whether the same
formalism can reproduce the magnetic moments.

\vskip 20pt
\subsection{Hamiltonian and vertex functions}

Although as a first approximation baryons can be treated as containing only
three quarks in the appropriate SU(6) wavefunction, the coupling to mesons
in the full Lagrangian means that there is a certain probability that, say,
a $\Delta$ will be made up of a bare nucleon plus a pion.  Since the coupling
is weak, it should be a good approximation to ignore contributions with two or
more mesons, and also to ignore the effect of the meson fields on the soliton
quark and $\chi$ fields.  The pion contributions have traditionally been
included in the cloudy bag model. Kaon
contributions however are also not negligible, and they also contribute to
mass splittings within the octet and decuplet, as was shown in
ref.~\cite{JMCD}.
On the other hand the heavy
mesons ($\eta'$ and the scalar) should have little effect, and they will
be ignored.  The interaction Lagrangian is then equivalent to one with a
non-linear realisation of the light mesons, and the formalism of the cloudy bag
can be adopted intact.  The reader is referred to ref.~\cite{T&T,CBM}
for details; the final baryon-meson interaction Hamiltonian is
\begin{equation}
H_I=\sum^8_{a=1} \int d^3\!k \sum_{A,B} A_0^\dagger B_0
[v_a^{AB}(\kvec) c_a(\kvec) +v_a^{BA}(\kvec)^*
c_a^\dagger(\kvec)]
\end{equation}
where $c_a^\dagger(\kvec)$ and $c_a(\kvec)$ are pseudoscalar octet meson
creation
and annihilation operators and $A_0^\dagger$ creates a bare (three-quark)
baryon
$A$. The vertex function $v_a^{AB}(\kvec)$ is given by
\begin{equation}
v_a^{AB}(\kvec)={ig_n\over f_\pi}\int {d^3\!r\over \left[(2\pi)^3
2\omega_a(k)\right]^\shalf} e^{i\kvec\cdot \rvec}
\langle A_0|:{ {\overline q} \lambda_a\gamma_5 q \over \chi^p}:
|B_0\rangle
\label{vertex}\end{equation}
where $|A_0\rangle $ is the  wavefunction of bare baryon $A$,
$\omega_a(k)=\sqrt{m_a^2+k^2}$ and $m_a=m_\pi$, $a=1-3$,
$m_\K$, $a=4-7$ or $m_\eta$, $a=8$.  In order to
calculate the matrix element in Eq.~\ref{vertex}
 correctly one would need a full quantum
wavefunction for the quarks and $\chi$ field in baryons $A$ and $B$ (a coherent
state, for instance).  Such sophistication would be wasted on a
perturbative calculation such as this.  There is in any case no problem for the
dominant, pion contributions, where $A$ and $B$ have the same hypercharge;
there the basic
quark-$\chi$ structure of each is the same and the matrix element can be
calculated in the mean-field approximation.  The problem arises with the
kaons.  These are however already less important than the pions, and
it is unlikely to cause major errors if we take the overlap of the $\chi$ part
of the wavefunctions to be unity here also.

With the quark spinor for the $i$th quark given by
\begin{equation}
q_i(\rvec)=\left({G_i(r)\atop i\sigbol\cdot\rvec\, F_i(r)}\right),
\label{spinor}\end{equation}
Eq.~\ref{vertex}  becomes
\begin{equation}
v_a^{AB}(\kvec)={-i\over \left[(2\pi)^3
2\omega_a(k)\right]^\shalf}u_a(k)\zeta_a
\langle A_{sf}|\sum_{i=1}^3 \lambda_a^{(i)} \sigbol^{(i)}\cdot \kvec
|B_{sf}\rangle.
\end{equation}
In the above  $|A_{sf}\rangle$ is the SU(6) spin-flavour wavefunction and
$u_a$ is a normalised formfactor
\begin{equation}
u_a(k)={3\int dr\, r^3 (G_iF_j+G_jF_i)/\chi^p\; j_1(kr)/kr \over
\int dr\, r^3 (G_iF_j+G_jF_i) /\chi^p},
\end{equation}
$i$ and $j$ are both non-strange for pions ($a=1-3$), one strange and
one non-strange for kaons ($a=4-7$), and the appropriate combination of
both strange and both non-strange for $\eta$ ($a=8$). In addition, the
factor $\zeta_a$ is given by
\begin{equation}
\zeta_a={4\pi g_n\over 3 f_\pi} \int dr\, r^3 (G_iF_j+G_jF_i) /\chi^p.
\end{equation}
Using the Wigner-Eckart theorem the vertex function can be written
\begin{eqnarray}
 v_a^{AB}({\bf k}) &=& -\frac i{[(2\pi)^32\omega_k]^{1/2}}
 \sum_{\nu} \zeta_{\nu}u_{\nu}(k)f_{\nu}^{AB} \sum_{q,i}
 \langle S_BS_{Bz}1q|S_AS_{Az} \rangle e_{qi}^*k_i \nonumber \\
 & & \mbox{\hspace{3.5cm}} \times \langle T_BT_{B3}T_{\nu}T_{\nu3}|T_AT_{A3}
 \rangle c_{\nu a}^* \;\;.
\end{eqnarray}
where
\begin{equation}
 f_{\nu}^{AB} = \sum_{\gamma} \left( \begin{array}{cc|c} d_B & 8 &
  d_{A,\gamma} \\ T_BY_B & T_{\nu}Y_{\nu} & I_AY_A \end{array} \right)
  R(d_A,d_B)_{\gamma} .
\label{vertex2}\end{equation}
and the reduced matrix elements are calculated to be~\cite{JMCD}
\[  R(8,8)_1 = 2\sqrt 5\;\;,\;\;\; R(8,8)_2 = 4\;\;,\;\;\;
  R(8,10) = -2\sqrt{10}\;\;, \]
\begin{equation}
  R(10,8) = -4\;\;,\;\;\; R(10,10) = 2\sqrt{10}  .
\end{equation}
The matrices
$[e]_{iq}$ and $[c]_{a\nu}$ are, respectively, the transformation matrix
between
the cartesian and spherical bases, and that between the basis in the $SU(3)$
regular representation and the basis in the isospin classification of $SU(3)$
multiplets.
They satisfy the orthogonality relations
\begin{equation}
   \sum_i e_{qi} e_{q'i}^* = \delta_{qq'} ,\;\;\;
   \sum_a c_{\nu a} c_{\nu' a}^* = \delta_{\nu\nu'}  .
\end{equation}

As an example of a meson contribution to an observable,
the mesonic self-energy of baryon $A$ is then given by
\begin{equation}
\Sigma_A=\sum_B \sum_a \int d^3\!k {v_a^{AB}(\kvec)
v_a^{*BA}(\kvec) \over \omega_a(k)
\bigl(M_A-M_B-\omega_a(k)\bigr)}
\end{equation}

In cloudy bag calculations \cite{T&T,CBM} it is customary to take $M_A$ and
$M_B$ as the {\it physical} baryon masses.  This is equivalent to summing some
self-energy diagrams to all orders, but without even including the first order
change in the quark wavefunctions due to the presence of the mesons.  It is
more consistent in a first-order perturbative calculation to use the
bare (or soliton) masses.

\section{Perturbative Contributions to the Magnetic Moment}

Th\'{e}berge and Thomas have calculated the magnetic moments of the octet
baryons in the cloudy bag model~\cite{T&T}. Our calculation mirrors theirs,
with two exceptions.  One is that we include kaon and $\eta$ loops.  The other
is that we calculated according to strict second-order perturbation theory,
so that our vertex functions are ``unrenormalised" and the terms
$\omega_{AB}=M_A-M_B$ which
occur in denominators are taken as bare (soliton), rather than physical, mass
differences.  Henceforth whenever reference is made to equations in their
paper, this difference should be borne in mind.

In terms of the  electromagnetic current operator $\hat{j}^{\mu}(x)$,
the magnetic moment of baryon $A$  defined as
\begin{equation}
  \mu^A \equiv
  \langle A,S_{Az}=S_A|\int d^3\!r\;{\bf r}\times \hat{{\bf j}}({\bf r}) \:
 |A,S_{Az}=S_A \rangle  ,
\label{magdef}\end{equation}
that is the magnetic moment is defined in terms of the spin-stretched state.
It is implied that the center-of-mass system of baryon $A$ is taken. Since it
is impossible to define a momentum eigenstate of a spatially localized soliton,
there arises a need to remove spurious center-of-mass motion. However,
center-of-mass corrections are not taken into account in the present work.

The  electromagnetic current
may be split into quark and meson terms. Accordingly, the magnetic moment
has three separate contributions, as shown diagrammatically in Fig.~1,
corresponding to the coupling of a photon to a bare baryon ($\mu_0^A$), to a
charged meson ``in flight" about a baryon
($\mu_\phi^A$), and to a baryon with a meson in flight ($\mu_Q^A$), and the
total is given by $\mu^A=Z_A(\mu_0^A+\mu_\phi^A+\mu_Q^A)$.
$Z_A$ is the wavefunction normalisation factor, given to the same order
in perturbation theory by
\begin{equation}
  Z_A^{-1} = 1+ \frac 1{12\pi^2} \sum_{B,\nu} (\zeta_{\nu}f^{AB}_{\nu})^2
    \int_0^{\infty} \frac{dk \; k^4 u_{\nu}^2(k)}
   {\omega_{\nu}(k)(\omega_{BA}+\omega_{\nu}(k))^2}  .
\label{norm}\end{equation}

Of course, within the framework of second order in perturbation theory
there is no need
to multiply the second order contributions  by a normalisation factor different
{}from one, but we chose to do so so that the coupling of a time-like photon
calculated in the same approximation does indeed give
a proton electric charge
of $+e$.  Convergence of the perturbation expansion is also improved.

We will deal with the diagrams in which the photon couples to a meson, and
those in which it couples to a quark (with or without a meson in flight)
separately in the next two sections.

\vspace{1.5cm}

\subsection{Meson-photon coupling}
\indent\indent
We start with examining the effect of a virtual pion coupling to a photon
corresponding to Fig.~1c.  This mirrors exactly the treatment in section 4.5 of
ref.~\cite{T&T}, with the caveats mentioned above, ending with the expression
\begin{equation}
 \mu_{\pi}^A = \frac { e}{18\pi^2} \sum_B  (\zeta_{\pi}f_{\pi}^{AB})^2
  s_A(B)t_A(B) \int_0^{\infty} dk \frac{k^4u_{\pi}^2(k)}{\omega_k^4}  .
\label{pimm}\end{equation}
The $\omega_{AB}$ vanish here since the nucleon and delta are built on the
same bare soliton.
The spin and isospin Clebsch-Gordan coefficients are summarized in
$s_A(B)$ which is given by
\begin{eqnarray}
  s_A(B) &=& \left\{ \begin{array}{cl} 1 &
  \mbox{\quad if \enskip} S_B=\frac12 \\ -\frac12 & \mbox{\quad if \enskip}
  S_B=\frac32 \end{array}\right\} \mbox{\ and \enskip} s_A=\half, \nonumber \\
  s_A(B)  &=& \left\{ \begin{array}{cl} \frac 12 &
  \mbox{\quad if \enskip} S_B=\frac12 \\ \frac 15 & \mbox{\quad if \enskip}
  S_B=\frac32 \end{array}\right\} \mbox{\ and \enskip} s_A=\textstyle{\frac
32},
\end{eqnarray}
and $t_A(B)$:
\begin{eqnarray}
  t_A(B) &=&  \langle T_BT_{B3}11|T_AT_{A3} \rangle^2
   -\langle T_BT_{B3}1{-1}|T_AT_{A3} \rangle^2  \\
  &=& \left\{ \begin{array}{cl} T_{A3}/T_A & \mbox{\quad if \enskip}
  T_B=T_A-1 \\ T_{A3}/[T_A(T_A+1)] & \mbox{\quad if \enskip} T_B=T_A
  \\ -T_{A3}/(T_A+1) & \mbox{\quad if \enskip} T_B=T_A+1
  \end{array}\right.  .
\end{eqnarray}

The coupling to kaons can be evaluated in a similar way.
Since the charged kaons and a combination of the neutral pion and $\eta$ form a
V-spin triplet, an expression for the kaon current very similar to Eq.~(4.32)
of ref.~\cite{T&T} is obtained:
\begin{equation}
  \hat{j}_\K^{\mu}({\bf r}) = -\frac{ie}2\int \frac{d^3\!kd^3\!k'}
  {(2\pi)^3\sqrt{\omega_k\omega_{k'}}} \; e^{-i({\bf k}'-{\bf k})\cdot{\bf r}}
  k^{\mu}\sum_{j',j=1}^2\epsilon_{j'j3} \hat{S}_\K({\bf k}'j',{\bf k}j,\mu)
   ,
\end{equation}
where $ \omega_k \equiv \omega_\K(k) $ and
\begin{equation}
 \hat{S}_\K({\bf k}'j',{\bf k}j,\mu) = (a_{j'}(-{\bf k}')
 +a_{j'}^{\dagger}({\bf k}'))(a_j({\bf k})-g_{\mu\mu}a_j^{\dagger}(-{\bf k}))
   .
\end{equation}
but where the labels on $a_1$ and $a_2$ correspond to V-spin, and in terms of
an octet of annihilation operators $c_a$, we have
$a_1\equiv c_4$ and $a_2\equiv c_5$.
The expectation values $S_\K^A$ take on a form identical to Eqns.(4.45) and
(4.46) of\ \cite{T&T}, written however in terms of the V-spin labelled
charged kaon vertex functions:
\begin{eqnarray}
 v_{Kj}^{AB}({\bf k}) &=& -\frac i{[(2\pi)^32\omega_k]^{1/2}} \, f_\K^{AB} \!\!
\! \sum_{q,i,r=\pm\shalf}\zeta_r u_r(k) \langle S_BS_{Bz}1q|S_AS_{Az} \rangle
  e_{qi}^*k_i \nonumber \\ & & \mbox{\hspace{3.5cm}}  \times
  \langle T_BT_{B3}{\textstyle \frac12}r|T_AT_{A3} \rangle e_{2r,j}^*  ,
\end{eqnarray}

Hence the magnetic moment due to the kaon-photon coupling acquires a form
similar to Eq.~\ref{pimm}:
\begin{equation}
 \mu_\K^A = \frac {e}{18\pi^2} \sum_{B,r=\pm\shalf}(\zeta_r f^{AB})^2
  s_A(B)t_A^r(B) \int_0^{\infty} dk\frac{ k^4u_r^2(k)(\omega_{BA}+2\omega_k) }
  { 2\omega_k^3(\omega_{BA}+\omega_k)^2 }.
\label{kmm}\end{equation}
where the sum over $B$ is over all baryons linked to $A$ by the emission of a
charged kaon, and $t_A^r(B)=(-1)^{r-\shalf}
\langle T_BT_{B3}{\textstyle \frac12} r  |T_AT_{A3} \rangle^2$.

The form is more complicated than the pion case because the formfactors for
positive and negatively charged kaons are not quite identical, and also because
the mass differences between solitons of different strangeness do not vanish.
If such differences were ignored the sum over $r$ could be
done analytically, and the results expressed in terms of ``V-scalar factors"
rather than the isoscalar factors $f^{AB}$ and ``$v_A(B)$" which depends on
the V-spin of A and B exactly as the pionic $t_A(B)$ does on the isospin.
\vspace{1.5cm}

\subsection{Quark-photon coupling}
\indent\indent
As the final perturbative contribution, we discuss the quark-photon coupling,
Figs.~1a and 1b, starting from the definition of the magnetic moment,
Eq.~\ref{magdef}.
The difference from the pion-only case is that whereas in that case, for
instance, the nucleon magnetic moments depended only on the bare nucleon, delta
and nucleon-delta transition magnetic moments, hyperon magnetic moments and
transition magnetic moments now enter.  Those bare moments are
listed in Table~2 of\ \cite{T&T} (see later
for comments on errors and sign conventions).

The contribution to the magnetic moment from photon-quark coupling is
\begin{equation}
  \mu_Q^A = \mu^{(0)}(A)+\mu^A_{QM},
\label{qmm1}\end{equation}
where $\mu^{(0)}(A)$ is the bare magnetic moment, corresponding to Fig.~1a, and
$\mu_{QM}^A$ is the magnetic moment corresponding to Fig.~1b in which
a photon couples to the quarks with a virtual meson in
flight.

Both parts of $\mu_Q^C$ involve the quantity
\begin{equation}
  \mu_{Q0}(A,B) \equiv \langle A_0,S_{Az}={\textstyle \frac12}|
  \hat{\mu}_{Qz} |B_0,S_{Bz}={\textstyle \frac12} \rangle  .
\end{equation}
The bare magnetic moment $\mu^{(0)}(A)$ in Eq.~\ref{qmm1} is related to
$\mu_{Q0}(A)$ as follows. For the octet members
\begin{equation}
   \mu^{(0)}(A) = \mu_{Q0}(A,A)  ,
\label{bmm}\end{equation}
while
\begin{equation}
   \mu^{(0)}(\Omega) = 3 \mu_{Q0}(\Omega,\Omega)  ,
\end{equation}
as a consequence of the definition of $\mu_{Q0}$ in terms of the $S_z=\frac12$
state rather than the spin-stretched state.

Using the form of the quark spinors in Eq.~\ref{spinor} we obtain
\begin{equation}
  \mu_{Q0}(A,B) =  \langle A_{sf},S_{Az}={\textstyle \frac12}|
  \sum_{i=1}^3 Q_i \sigma_z^{(i)}\; \frac{8\pi}3 \int_0^{\infty}\! dr\,
  r^3G_iF_i |B_{sf},S_{Bz}={\textstyle \frac12}\rangle ;
\end{equation}
there are no ambiguities in the choice of wavefunction since
the magnetic moment operator does not change flavour.
Note that
\begin{equation}
  \mu_{Q0}(A,B) = \mu_{Q0}(B,A).
\end{equation}
The $\mu_{Q0}(A,B)$, are expressed in terms of
\begin{equation}
 \mu_f \equiv \frac{8\pi e}3 \int_0^{\infty} dr\;r^3G_fF_f   ,
\end{equation}
where the quark flavour $f$ is either $n$, non-strange, or $s$, strange.
They are given in Table 2 of ref.~\cite{T&T}.
However it should be noted that according to the sign convention used there,
the signs of $\mu_{Q0}(\Sigma^-,\Sigma^{*-})$ and $\mu_{Q0}(\Xi^-,\Xi^{*-})$
are
given wrongly.
Furthermore, to obtain agreement with the
the sign conventions of most tables of isoscalar factors\ \cite{pdt},
all the $\mu_{Q0}(\Sigma,\Sigma^*)$, $\mu_{Q0}(\Sigma,\Lambda)$ and
$\mu_{Q0}(\Xi,\Xi^*)$  should be multiplied by $(-1)$.  The set is completed
by  $\mu_{Q0}(\Omega,\Omega)=-\altext 1 3\mu_s$.

The second order piece $\mu_{QM}^A$ is given by
\begin{equation}
  \mu_{QM}^A = \sum_{a,B,C} \int d^3\!k \frac{v_a^{AB}({\bf k})}
  {(\omega_{BA}+\omega_a(k))} \langle B_0|\hat{\mu}_{Qz}|C_0 \rangle
  \frac{v_a^{*AC}({\bf k})}{(\omega_{CA}+\omega_a(k))}
\end{equation}
(equation 4.72 of\ \cite{T&T}, but allowing kaon and $\eta$ loops).
Substituting the vertex functions, Eq.~\ref{vertex2}, and manipulating the
Clebsch-Gordan coefficients yields
\begin{eqnarray}
 \mu_{QM}^A &=& \frac{ e}{36\pi^2}\sum_{\nu,B,C}\zeta_{\nu}^2f_{\nu}^{AB}
  f_{\nu}^{AC}S_{BC} \int_0^{\infty} \frac{dk\;k^4u_{\nu}^2(k)}
  {\omega_{\nu}(k)(\omega_{BA}+\omega_{\nu}(k))^2} \times \nonumber \\
& &
\sum_{T_{B3}} \langle T_BT_{B3}T_{\nu}T_{\nu3}|T_AT_{A3} \rangle
 \langle T_CT_{C3}T_{\nu}T_{\nu3}|T_AT_{A3} \rangle \mu_{Q0}(B,C) \;,
\label{qmm2}\end{eqnarray}
where the relation $\omega_{BA}=\omega_{CA}$ has been used, and $S_{BC}$ is
\begin{center}
\begin{tabular}{|c||c|c|c|} \hline
 $_{S_A} \;\; ^{S_B}$ & $=S_C=\frac12$ & $ \;\;\; \ne S_C \;\;\;$
 & $=S_C=\frac32$ \\ \hline\hline
 $\frac12$ & $-1$ & $-2$ & $5$ \\ \hline
 $\frac32$ & $3$ & $-\frac35\sqrt{10}$ & $\frac{33}5$ \\ \hline
\end{tabular}
\end{center}

As was mentioned in sect.~(2.1), appropriate combinations of non-strange and
strange form factors, $u_n(k)$ and $u_s(k)$, must be taken for the
$\eta$-meson, since it couples to both types of quarks:
\begin{equation}
 \zeta_{\eta}u_{\eta}(k) = C_n^{AB}\zeta_nu_n(k)+C_s^{AB}\zeta_su_s(k)  ,
\end{equation}
where the coefficients $C_n^{AB}$ and $C_s^{AB}$ depend on the transition
being considered. They are given in Table~1 of ref.~\cite{JMCD}.

\vspace{1.5cm}
\subsection{Summary of the formalism}
\indent\indent
To summarize, we write down a formula for the
perturbative corrections to the magnetic moment of baryon $A$
\begin{eqnarray}
   \mu^{(1)}(A)& =& Z_A [ {\mu}_{\pi}^A+{\mu}_\K^A+{\mu}_{Q\pi}^A
 +{\mu}_{Q\K}^A+{\mu}_{Q\eta}^A ] \nonumber\\
& = & \frac{Z_A e}{36\pi^2} \left[ \; D^A_{\pi} I^A_{\pi} +
 \sum_{B} D^{AB}_\K I^{AB}_\K + (D^A_Q(\pi n)\mu_n+D^A_Q(\pi s)\mu_s)
I^A_{Q\pi}  \right. + \label{ptmm}\\
&  & \left. \sum_{B} [(D^{AB}_Q(K n)\mu_n+D^{AB}_Q(K s)\mu_s) I^{AB}_{Q\K}
 + (D^{AB}_Q(\eta n)\mu_n+D^{AB}_Q(\eta s)\mu_s) I^{AB}_{Q\eta}]
 \; \right] \;, \nonumber
\end{eqnarray}
where taken out and
\[ I^A_{\pi} = \zeta_n^2 \int_0^{\infty}dk \frac{2k^4u_n^2(k)}
  {\omega_{\pi}^4(k)} \;, \;\; I^{AB}_\K = \zeta_{ns}^2 \int_0^{\infty}dk
  \frac{k^4u_{ns}^2(k)(\omega_{BA}+2\omega_\K(k))}
  {\omega_\K^3(k)(\omega_{BA}+\omega_\K(k))^2} \;, \]
\[ I^A_{Q\pi} = \zeta_n^2 \int_0^{\infty}dk \frac{k^4u_n^2(k)}
  {\omega_{\pi}^3(k)} \;, \;\;  I^{AB}_{Q\K} = \zeta_{ns}^2 \int_0^{\infty}
  \frac{dk\;k^4u_{ns}^2(k)} {\omega_\K(k)(\omega_{BA}+\omega_\K(k))^2} \;, \]
\begin{equation}
  I^{AB}_{Q\eta} = \int_0^{\infty} dk \frac{k^4}{\omega_{\eta}^3(k)}
  (D^{AB}_n\zeta_n u_n(k) + D^{AB}_s\zeta_s u_s(k))^2  .
\end{equation}
For ease of notation, the dependences of the meson masses on charge and of
the form factors on hypercharge are not explicitly shown, though taken
into consideration in actual calculations.  In particular,
transitions induced by kaons involve intermediate states $B$ which have
different values of hypercharge, so that $\omega_{BA}$ can vary over a range of
a few hundred MeV. For a check of numerical calculation it is useful to
examine the coefficients
\begin{equation}
  D^A_\K \equiv \sum_{B} D^{AB}_\K \;,\;\;
  D^A_Q (K f) \equiv \sum_{B} D^{AB}_Q (K f) \;,\;\;
  D^A_Q (\eta f) \equiv \sum_{B} D^{AB}_Q (\eta f)
\end{equation}
and $D_{\pi}^A,\;D_{Q}^A(\pi f)$, which we tabulate in Table~1.

Some relationships between $D_{\pi}$ and $D_\K$ are worth pointing out. As was
noted in sect.~3.1 there is a close analogy between isospin in the
pion-photon coupling and V-spin in the kaon-photon coupling. This leads to the
equality
\begin{equation}
 (D_{\pi}^p,D_{\pi}^n)=(D_\K^{\Sigma^+},D_\K^{\Xi^0})  ,
\label{vspin1}\end{equation}
which may be viewed as the relation between the isospin-$\frac 12$ pair
$(p,n)$ and the V-spin-$\frac 12$ pair $(\Sigma^+,\Xi^0)$. Similarly the
following relations are observed:
\begin{equation}
 (D_{\pi}^{\Sigma^+},D_{\pi}^{\Sigma^-})=(D_\K^p,D_\K^{\Xi^-}) \;, \;\;
 (D_{\pi}^{\Xi^0},D_{\pi}^{\Xi^-})=(D_\K^n,D_\K^{\Sigma^-})  .
\label{vspin2}\end{equation}
Note, however, that the $\Lambda$ and $\Sigma^0$ are not eigenstates of V-spin
but the appropriate combinations of $V=0$ and $V=1$ states, and therefore no
straightforward relations between $D_{\pi}$ and $D_\K$ exist in these cases.

The final result for the magnetic moment is
\begin{equation}
   \mu(A) = Z_A \mu^{(0)}(A) + \mu^{(1)}(A).
\end{equation}
One of the goals of the present work is to reveal the way in which the
predictions of the model are influenced by the effects of kaons and eta.
In fact, it has been shown\ \cite{JMCD} that the inclusion of those effects
are crucial for reproducing the observed mass spectrum of the octet and
decuplet baryons. It is therefore useful to study the two classes of magnetic
moment:
\begin{eqnarray}
   \mu_{pion}(A) &=& Z_{pion}^A
     (\mu^{(0)}(A)+{\mu}_{\pi}^A+{\mu}_{Q\pi}^A)  \label{mupi}\\
   \mu_{all}(A) &=& Z_{all}^A
     (\mu^{(0)}(A) +{\mu}_{\pi}^A+{\mu}_\K^A+{\mu}_{Q\pi}^A
     +{\mu}_{Q\K}^A+{\mu}_{Q\eta}^A)    , \label{muall}
\end{eqnarray}
where $Z_{pion}^A$ and  $Z_{all}^A$ correspond to
the cases in which intermediate states in Eq.~\ref{norm} include pions alone
and all types of mesons, respectively.

\section{Results and Discussion}
\subsection{CDM results and input parameters}

Armed with the formalism developed in the preceding two chapters, we
now investigate numerically how the magnetic moments of the octet
baryons and $\Omega$ can be described in our theoretical framework.
The present model is successful in reproducing
the mass spectrum of low-lying baryons\ \cite{JMCD}, and we wish to calculate
the magnetic moments with the same parameters.  It is also of interest to
determine the sensitivity to the model parameters, which are as follows.
The parameter $p$ is the power of the $\chi$ field in Eq.~\ref{lagrange},
which is taken to be one or two. $\beta$ is the dimensionless
coupling constant of a non-strange quark to the $\chi$ field. $c_g$ of
Eq.~\ref{cg} is the ratio of the non-strange to strange coupling strengths.
The experimental values of $f_{\pi}$ and $f_\K$ give $c_g = 1.44$.
It has been suggested\ \cite{JMCD} that, if the approximation of holding the
scalar, isoscalar at their vacuum value is lifted and they are allowed to
interact with the quark and $\chi$ fields, the effective coupling ratio inside
the soliton is enhanced  to $c_g=1.58$. We therefore study this ratio too.

As was described in sect.~2, the $\chi$-field mass $M_{\chi}$ is determined by
fitting the r.m.s. quark radius of the nucleon to its isoscalar charge radius
$r_I$. Once determined, $M_{\chi}$ fixes the overall  energy scale of the
theory prescribed by the solitonic Lagrangian, Eq.~\ref{lagrange}.
The literature shows a
considerable uncertainty in the  experimentally extracted value of the charge
radius of the proton\ \cite{radp} as well as that of the neutron\ \cite{radn},
and thus $r_I$ ranges over  $0.74 < r_I < 0.85$ fm. Furthermore it has been
shown\ \cite{Leech} that the correction of spurious center-of-mass  motion
associated with a localized soliton reduces the isoscalar radius $r_I$  by a
factor of some 10\%.  In view of these uncertainties in $r_I$, we have
considered a range of values from 0.72 (used in ref.~\cite{JMCD}) to 1~fm.  It
turns out the ratios of magnetic moments  are very insensitive to this
parameter.  It should be stressed that $M_{\chi}$ is
determined once $p,\;\beta$ and $r_I$ are specified and is therefore not an
additional input parameter.

There is one source of ambiguity still left, namely the
scale of the magnetic moment.  The non-relativistic reduction gives results
expressed in terms of the nuclear magneton $\mu_\N = e/2m_\N$, but it is not
obvious what
value of the nucleon mass $m_\N$ to take. (This problem is common to many
models\ \cite{Schwesinger})  Ideally, $m_\N$ ought to be calculated
within the framework of our model. However, former work on the same model
has shown\ \cite{JMCD} that the best-fit input parameters, set (II) in Table~5
with $r_I=0.72$ fm, yield a nucleon mass of 1227 MeV, about 30\% above
the experimental value. Although the level scheme of mass of the
octet and decuplet baryons is reproduced reasonably well, it should be kept in
mind that there is an uncertainty of up to 30\% in the absolute energy scale.
We use the experimental value of $\mu_\N$ to calculate what we term the
`absolute magnetic moment', which is subject to uncertainty originating from
the ambiguity in scale.
The absolute value of the proton magnetic moment varies from 1.96 at
$r_I=0.72$~fm to $2.61$ at $r_I=1.0$~fm.
This uncertainty may be removed by setting the absolute magnetic moment of one
specific baryon, for example the proton, to the corresponding experimental
value
and by scaling the absolute magnetic moments of other baryons accordingly.
The results thus obtained are called `scaled magnetic moments'.
In what follows we will discuss the implications of the present theory based
primarily on the scaled magnetic moments.

As a representative case of the $p=1$ model, we select $r_I=0.90$ fm,
$c_g=1.44$ and $\beta=0.028$.
In Table~2 we show the various components of magnetic moment defined in
Eqns.~\ref{bmm} and \ref{ptmm}, the
comparison between the effects of pions alone and those of all mesons,
{\it i.e.} $\mu_{pion},\; \mu_{all},\; Z_{pion},\; Z_{all}$ in Eqns.~\ref{mupi}
and \ref{muall}, and the experimental moments $\mu_{exp}$ with their errors.
All
magnetic  moments are given in terms of absolute values, except scaled magnetic
moments  $\tilde{\mu}_{pion}$ and $\tilde{\mu}_{all}$.

Fig~2 shows the scaled magnetic moments for these parameters (ii), and then
with $r_I=0.7$~fm and other parameters as before (iii), $\beta=0.1$ (iv) and
$c_g=1.58$ (v).  Finally, the results for the $p=2$ model, with $r_I=0.90$ fm,
$c_g=1.44$ and $\beta=0.00245$ (vi) are shown.

The most striking feature of the results is that the colour-dielectric model
does a remarkably good job of describing the magnetic moments, with a maximum
deviation of 0.27 and an average deviation of only 0.11 nuclear magnetons.
It can also be seen that changing $r_I$ and $\beta$ and $c_g$ has essentially
no effect on the scaled magnetic moments.
As was discussed in sect.~2, the bare level of our theory does not
depend on the parameter $\beta$ once the overall energy scale is fixed.
It is evident that the meson masses hardly spoil the scaling property: the pion
masses are small on the typical energy scale of low-energy phenomena.
Though the kaon masses are much larger, the values of the kaon contributions
are too small to spoil the scaling.

However fit for the $p=2$ case
is distinctly worse.  In particular there is inadequate SU(3) breaking, as
can be seen by the reduced $\mu_p-\mu_{\Sigma^+}-|\mu_\Omega|$,
$\mu_n-\mu_{\Xi^0}$ and $\mu_{\Sigma^-}-\mu_{\Xi^-}$ splittings.  (All are
zero in an SU(3) symmetric model.)  This agrees with the results of
ref.~\cite{JMCD}, where the octet and decuplet mass splittings were found to be
unsatisfactorily  small in the $p=2$ model.

A notable feature of the results in Table~6 is the small influence on the
magnetic moments of the effects of kaon and $\eta$ mesons.
Columns (ii) to (iv) of Fig.~3, show the scaled magnetic moments without
mesons, with pions, and with kaons and the $\eta$ in addition.  With pions
alone the SU(3) splittings referred to above are almost perfect; including
kaons
decreases them slightly.  The contributions of all the diagrams involving kaons
are small, and  they are almost offset by the increased  value of the inverse
normalisation constant $Z^{-1}$. Similar results have been found in the
CBM\ \cite{cbm kaon}.The origin of these small contributions  can
be traced back to small values of the integrals in  Eq.~\ref{kmm}. A typical
denominator of the integrand contains at least the  third power of the meson
mass, which, together with the form factor, greatly  suppresses the integral
value. In contrast, the self-energy contribution of  mesons involve the
second-power of the meson mass, so that the  integral is not expected to be
reduced so much. In fact, the inclusion of  kaons and $\eta$ is essential for
describing the mass spectrum of  low-lying baryons\ \cite{JMCD}.

The $\Omega$ contains three strange and no non-strange valence quarks,
so that $\mu_\pi = 0$ and $Z_{pion} = 1$.
The successful reproduction, as can be seen in Fig.~3, of the $\Omega$
magnetic moment at the pion level without the inclusion of kaon and $\eta$
effects can thus be regarded as evidence the bare solitons are a good starting
point for a description of baryons.
We should particularly recall that our model provides a natural
mechanism for determining the ratio between the non-strange and strange
coupling
strengths. It is an experimentally fixed parameter in the hidden-symmetry
realization of an approximate $SU(3)_R \times SU(3)_L$ symmetry of our model
Lagrangian.

\subsection{Comparison with other Models}

Direct comparison with the results of ref.~\cite{T&T} for the  cloudy bag model
is complicated by the use
there of the physical masses, renormalised coupling constants and the exclusion
of kaons, as well as the inclusion of estimates of the centre-of-mass
corrections.  Thus we have repeated our calculation to produce results
for the CBM treated in the same manner as the CDM.
These are shown in the last two columns of Fig.~3.  (The bare values are
indistinguishable, so that  column (ii) could be either model).  Pions only are
included in (v), while kaons and the $\eta$ are added in (vi).  The strange
quark mass has been taken as 144~MeV and the bag radius as 1~fm, the preferred
parameters of Th\'eberge and Thomas\ \cite{T&T}. There is a strong resemblance
between the two models, as might be expected from the common methods used. The
fit in the CBM is poorer, but would probably be improved  by a larger strange
quark mass.  There too, kaons play very little role.

It is worth noting that in the
cloudy bag model, the inclusion of one-gluon exchange improves overall
agreement and is vital for recovering the correct order of
$\mu(\Lambda)$ and $\mu(\Xi^-)$\ \cite{level crossing}.
Furthermore, center-of-mass corrections can in principle be carried out
unambiguously by projecting a soliton on to an eigenstate of momentum. By
eliminating spurious center-of-mass motion, a significant improvement of
nucleon properties has been achieved in two-flavor versions of the
color-dielectric model\ \cite{Leech,Leech2}.
This is an advantage of a model which consists entirely of dynamical fields.
Although the same technique has been applied to the MIT bag model\
\cite{MITcm},
ambiguities associated with the static bag are unavoidable.

The magnetic moments of baryons have now been calculated in many models, most
of which claim to give a reasonable fit.  To the extent that most (but not all)
correctly reproduce the order of the moments (except
$\mu(\Lambda)$ and $\mu(\Xi^-)$) this may be true.  However all models are
certainly not equal in the quality of fit, as can be seen from Fig.~4, where
published results from a number of models are shown (all scaled to give the
experimental proton moment).  As a rough guide, $\chi^2$ values for all the
fits have been calculated, excluding the $\Omega$ and in each case allowing the
overall scale to vary to minimise $\chi^2$.  The theoretical values have been
assigned an arbitrary error  of
0.11 so that the $\chi^2$ per degree of freedom  is about 1
(1.1) for the CDM.

First, it can be seen from column (iii) that lattice calculations
(Leinweber\ \cite{lattice magmom}) now rival most dynamical models.
With $\chi^2=1.5$ the results are only slightly worse that the CDM.
A non-relativistic model with pion-exchange
currents (Wagner {\it et al}\ \cite{Wagner})  does not give sufficient SU(3)
breaking.  It is shown in column (iv), and with $\chi^2=2.2$ is not as good as
a simple additive quark model with $\mu_s=0.6\mu_n$
($\chi^2=1.6$, not shown).  These models however all do very much better that
the remaining class of models, including the Skyrme model, the NJL model and
the chiral bag model, all of which are based on the hedgehog. Column (viii) is
the SU(3) Skyrme model calculation of Park, Schechter and Weigel\ \cite{Park1},
($\chi^2=8.3$) which clearly has inadequate SU(3) breaking.
Park and Weigel\ \cite{Park} also included vector mesons which actually
reverses the order of two of the SU(3) splittings ($\chi^2=9.9$, not shown).
The bound state treatment of the Skyrme model, (vii) (Oh, Min and Rho\
\cite{Oh}; $\chi^2=6.8$) has the same failing.  At zero bag radius the chiral
bag model (Hong and Brown\ \cite{Hong};  $\chi^2=11.2$) resembles the Skyrme
model; at $r_{\rm bag}=1$~fm the fit is  a little better ($\chi^2=10.1$;
column (vi)) but
the $\Sigma^-$ is particularly bad, falling below the $\Lambda$.  The NJL model
(Kim {\it et al}\ \cite{Kim}; $\chi^2=6.0$, not shown) is the best of the
projected hedgehogs.  It is clear that all these models have problems in the
treatment of kaon fields.  In the SU(3) Skyrme model, for instance, kaons
appear when the embedded SU(2) pionic hedgehog is rotated in flavour space.  As
a result, the kaons have a profile governed by the pion mass.  Schwesinger and
Weigel\ \cite{Schwesinger} have performed  calculations in the Skyrme model in
which they have allowed the kaon profile to vary (the ``slow rotor approach"),
 with much improved results.
(Column (v) shows their results for the usual Skyrme term; $\chi^2=2.5$
dominated by the poor neutron moment).

Finally, it is worth comparing our results with those of Jenkins {\it et al} in
chiral perturbation theory\ \cite{Jenkins}.
 Expanding meson loop diagrams in powers of the
quark masses, they obtain predictions for the SU(3) splittings which are all
too large by factors of 2.5 to 4.8 if kaons are included, and poor results
still without kaons.  In contrast, the worst in the CDM with
kaons is the $\mu_{\Sigma^-}-\mu_{\Xi^-}$ splitting, at 60\% of its
experimental value, while with pions alone this rises to 80\%, with the other
two near perfect.  On the other hand the relations which hold to
$O({m_q}^\shalf)$ are also extremely well satisfied in the CDM:
$$\mu_{\Sigma^+}=-2\mu_\Lambda-\mu_{\Sigma^-},$$
$$\mu_{\Xi^0}+\mu_{\Xi^-}+\mu_n=2\mu_\Lambda-\mu_p;$$
the first is satisfied to within 2\% and the second to within 0.3\% with or
without kaons.  Thus it is clear that the fact that the experimental moments
approximately satisfy these relations is not related to the convergence or
otherwise of the  perturbation
expansion, and that loop diagrams treated to all orders agree much better
with experiment that the first term in their chiral expansion.  Similar
behaviour
has also been noted in the context of baryon self energies.\ \cite{Birse &
Stuckey}

\section{Conclusion}

We have calculated the octet and $\Omega$ magnetic moments in the
colour-dielectric model, in which pion and kaon fields have been included
perturbatively.  Good agreement with experimental values are obtained
for the scaled magnetic moments, in an approach in which $\mu_p$ is effectively
the only free parameter.  In particular the strength of SU(3) breaking is
governed entirely by the pion and kaon masses and decay constants. However the
absolute values are subject to some uncertainty.  This work confirms that
models based on the perturbative inclusion of mesons give very much better
results for the magnetic moments  than those based on hedgehogs, and while the
colour-dielectric model and  the cloudy bag model give very similar results,
the former has the advantage of being fully dynamical.
Finally the comparison with chiral perturbation theory suggests that in the
work which has been done so far, the expansion in powers of the strange quark
mass has not converged. \

\newpage

\def\NP#1{Nucl.\ Phys.\ {\bf  #1}}
\def\PR#1{Phys.\ Rev.\ {\bf  #1}}
\def\PRL#1{Phys.\ Rev.\ Lett.\ {\bf  #1}}
\def\PL#1{Phys.\ Lett.\ {\bf  #1}}
\def\ibid#1{{\it ibid}\hbox{\ {\bf  #1}}}

\newpage
\renewcommand{\arraystretch}{2.5}
\begin{table}[h] \caption{$D^A$}
\begin{center}
\begin{scriptsize}
\begin{tabular}{|c||c|c|c|c|c|c|c|c|} \hline
 $ A $ & $\;\;\;\;D_{\pi}\;\;\;$ & $\;\;\;D_\K\;\;\;$ & $D_Q(\pi n)$
 & $D_Q(\pi s)$ & $D_Q(K n)$ & $D_Q(K s)$ & $D_Q(\eta n)$ & $D_Q(\eta s)$ \\
 \hline\hline
 $p$        & 22  & 8   & 145           & 0
  & 36  & 2              & -1             & 0 \\ \hline
 $n$        & -22 & -2  & -100          & 0
  & -24 & 2              & $\frac23$      & 0 \\ \hline
 $\Sigma^+$ & 8   & 22  & $\frac{88}3$  & $\frac{20}3$
  & 102 & $\frac{40}3$   & $\frac{256}9$  & $\frac{20}9$ \\ \hline
 $\Sigma^-$ & -8  & 2   & $-\frac{56}3$ & $\frac{20}3$
  & -50 & $\frac{40}3$   & $-\frac{128}9$ & $\frac{20}9$ \\ \hline
 $\Lambda$  & 0   & -12 & 0             & -36
  & -2  & -24            & 0              & $\frac43$ \\ \hline
 $\Xi^0$    & -2  & -22 & 0             & -12
  & -32 & $-\frac{154}3$ & $-\frac{94}9$  & $-\frac{140}9$ \\ \hline
 $\Xi^-$    & 2   & -8  & 1             & -12
  & 12  & $-\frac{154}3$ & $\frac{47}9$   & $-\frac{140}9$ \\ \hline
 $\Omega$   & 0   & -18  & 0             & 0
  & -6  & -72            & 0              & -44 \\ \hline
\end{tabular}
\end{scriptsize}
\end{center} \end{table}

\vspace{2.5cm}
\noindent
The coefficients $D^A$ defined in Eq.~\ref{ptmm}. Note the relations
between $D_{\pi}$ and $D_\K$, Eqns.~\ref{vspin1} and \ref{vspin2}.

\newpage

\def\nl{\\ \hline}
\renewcommand{\arraystretch}{2}

\begin{table}[h] \caption{Various components of the magnetic moment}
\begin{center}
\begin{scriptsize}
\catcode`?=\active \def?{$-$}
\begin{tabular}{|c||r|r|r|r|r|r|r|r|} \hline
& $p\;$ &  $n\;$ &  $\Sigma^+\;$ & $\Sigma^-\;$ & $\Lambda\;$  & $\Xi^0\;$
& $\Xi^-\;$  & $\Omega\;$   \nl
$\mu^{(0)}$   &  1.99 & ?1.33 &  1.96 & ?0.69 & ?0.58 & ?1.22 & ?0.56 &
?1.76\nl
$\mu_{\pi}$  &  0.79 & ?0.79 &  0.29 & ?0.29 &  0.00 & ?0.07 &  0.07 &  0.00\nl
$\mu_\K$      &  0.04 & ?0.01 &  0.09 &  0.01 & ?0.05 & ?0.09 & ?0.04 &
?0.08\nl
$\mu_{Q\pi}$ &  0.96 & ?0.66 &  0.24 & ?0.09 & ?0.21 & ?0.07 & ?0.06 &  0.00\nl
$\mu_{Q\K}$   &  0.05 & ?0.03 &  0.17 & ?0.06 & ?0.03 & ?0.12 & ?0.04 &
?0.12\nl
$\mu_{Q\eta}$&  0.00 &  0.00 &  0.04 & ?0.01 &  0.00 & ?0.03 & ?0.01 & ?0.04\nl
$Z_{pion}^{-1}$ & 1.57  & 1.57  & 1.20  & 1.20  & 1.36  & 1.09  & 1.09
&1.00\nl
$Z_{all}^{-1}$ & 1.61  & 1.61  & 1.32  & 1.32  & 1.45  & 1.24  & 1.24  &
1.13\nl
$\mu_{pion}$ &  2.38 & ?1.78 &  2.07 & ?0.89 & ?0.58 & ?1.25 & ?0.50 & ?1.76
\nl
$\mu_{all}$ &  2.38 & ?1.76 &  2.11 & ?0.86 & ?0.60 & ?1.29 & ?0.52 & ?1.78
\nl
$\tilde{\mu}_{pion}$ &  2.79 & ?2.08 &  2.42 & ?1.04 & ?0.68 & ?1.46 & ?0.59
 & ?2.06  \nl
$\tilde{\mu}_{all}$ &  2.79 & ?2.07 &  2.47 & ?1.01 & ?0.71 & ?1.52 & ?0.61
 & ?2.09  \nl
$\mu_{exp}$&  2.79 & ?1.91 &  2.42 & ?1.16 & ?0.61 & ?1.25 & ?0.65  & ?1.94
\vspace{-7pt}\\
& $ (\pm0.00)$ & $(\pm0.00)$ & $ (\pm0.05)$  & $(\pm0.03)$ & $(\pm0.00)$
& $(\pm0.01)$ & $(\pm0.00)$  & $(\pm0.22)$ \nl

\end{tabular}

\end{scriptsize}
\end{center} \end{table}

\vspace{1cm}
\noindent
The breakdown of the magnetic moments of the octet baryons and $\Omega$
which are calculated for the parameter set $r_I=0.90$ fm, $\beta=0.028$,
$c_g=1.44$, $p=1$. See Eqns.~\ref{pimm}, \ref{kmm}, \ref{bmm},
\ref{qmm2} and \ref{ptmm} for the
definitions of the various components of magnetic moment. $\tilde{\mu}_{pion}$
and $\tilde{\mu}_{all}$ are scaled magnetic moments while the rest are
absolute magnetic moments.
\newpage
{\Large\bf Figure captions}                  \\

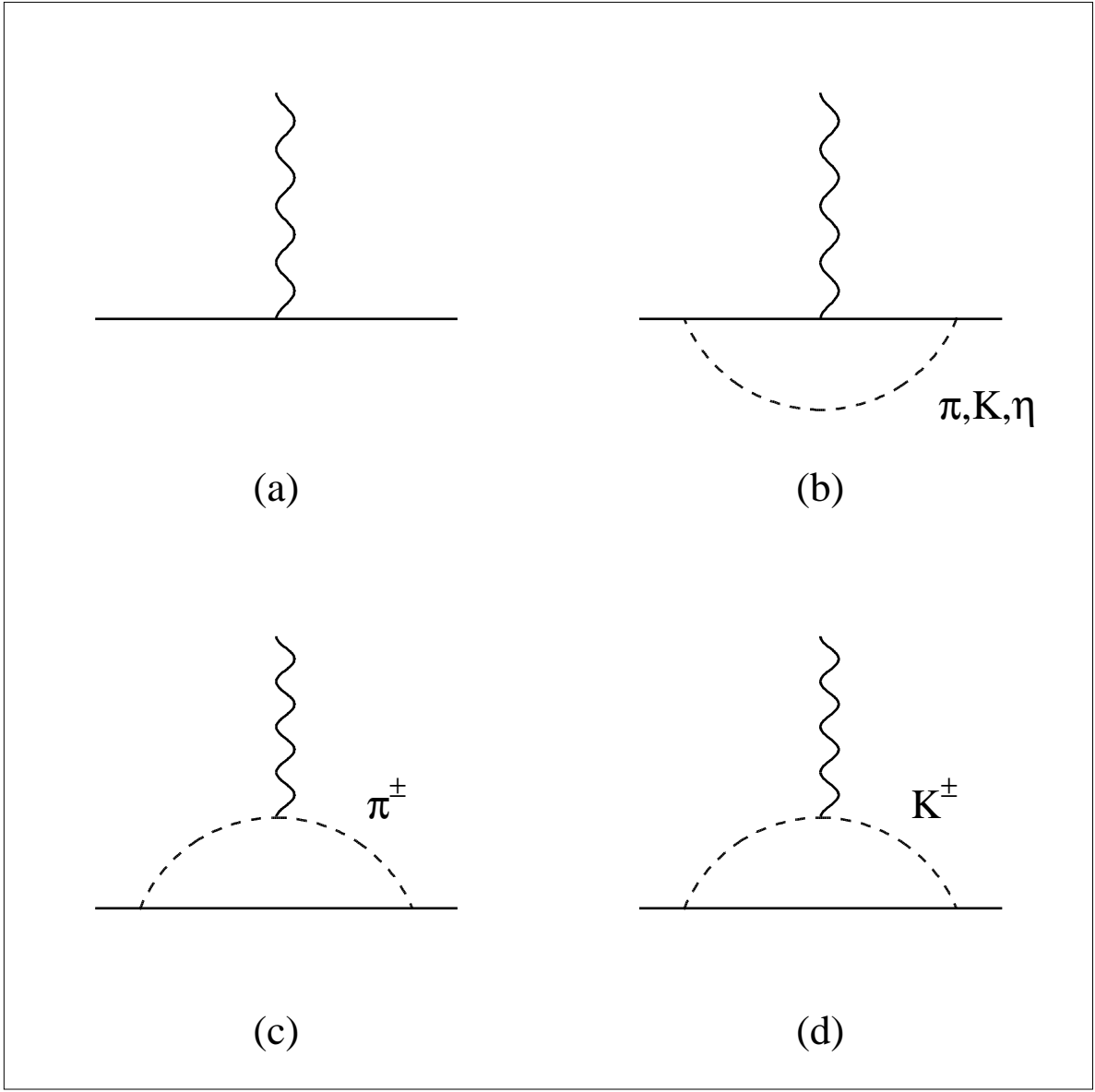
{\bf Fig.~1} \quad Four types of contributions to the magnetic moment:
(a) bare, \ (b) quark-photon coupling, \ (c) charged-pion-photon coupling, \
(d) charged-kaon-photon coupling. The solid, dashed and wavy lines represent
a bare baryon, a virtual meson dressing it and a probing photon, respectively.
\\

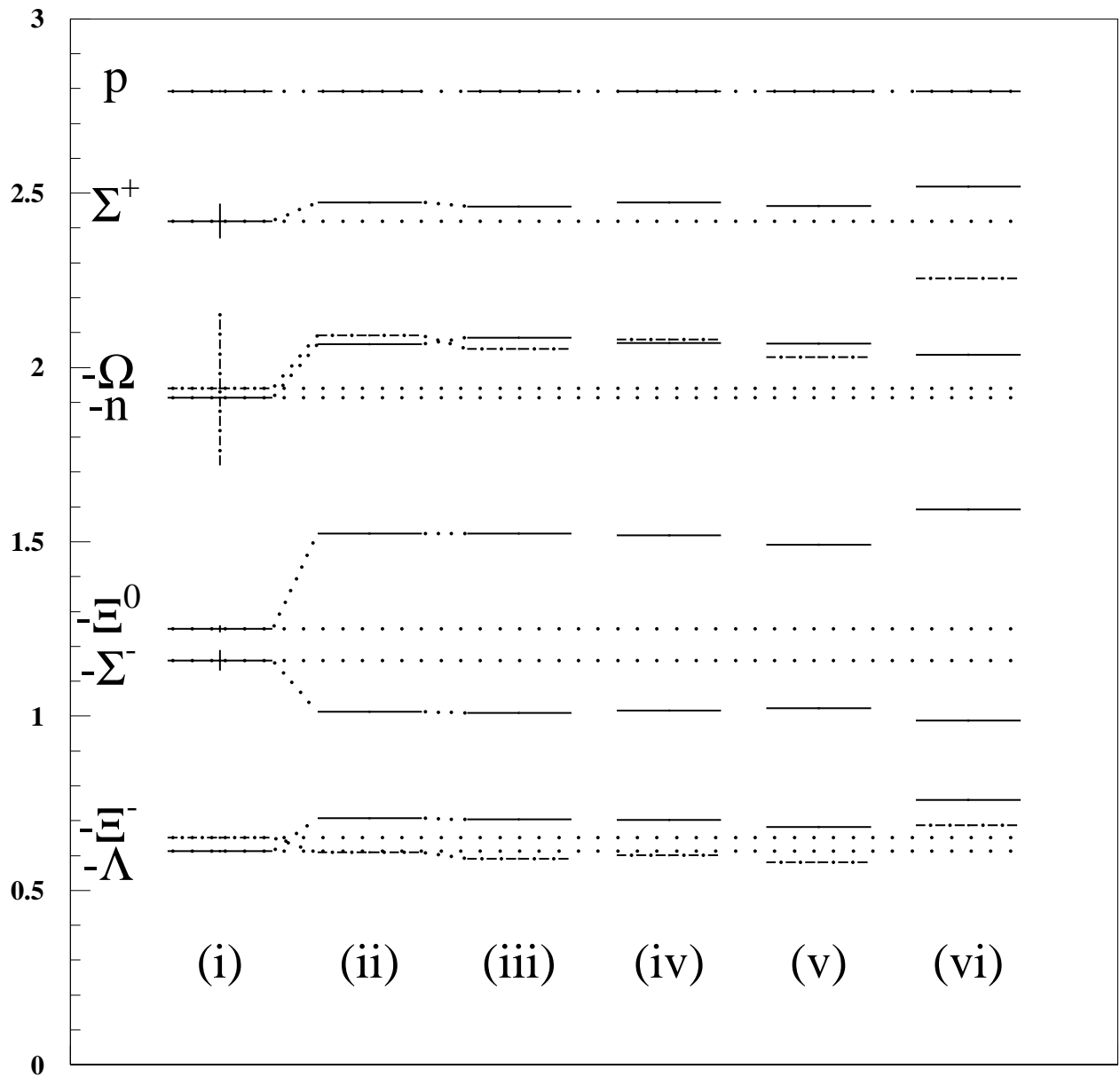
{\bf Fig.~2} \quad The scaled magnetic moments calculated for various input
parameters. \\ (i) Experimental magnetic moments; \\ (ii) $p=1$, $c_g=1.44$,
$r_I=0.9$~fm and $\beta=0.028$ (hence $M_\chi=1883$~MeV);\\
(iii) as (ii) but $r_I=0.72$~fm ($M_\chi=2354$~MeV);\\
(iv) as (ii) but $\beta=0.1$ ($M_\chi=805$~MeV);\\
(v) as (ii) but $c_g=1.58$;\\
(vi) $p=2$, $c_g=1.44$, $r_I=0.9$~fm and $\beta=0.00254$ ($M_\chi=1676$~MeV);\\

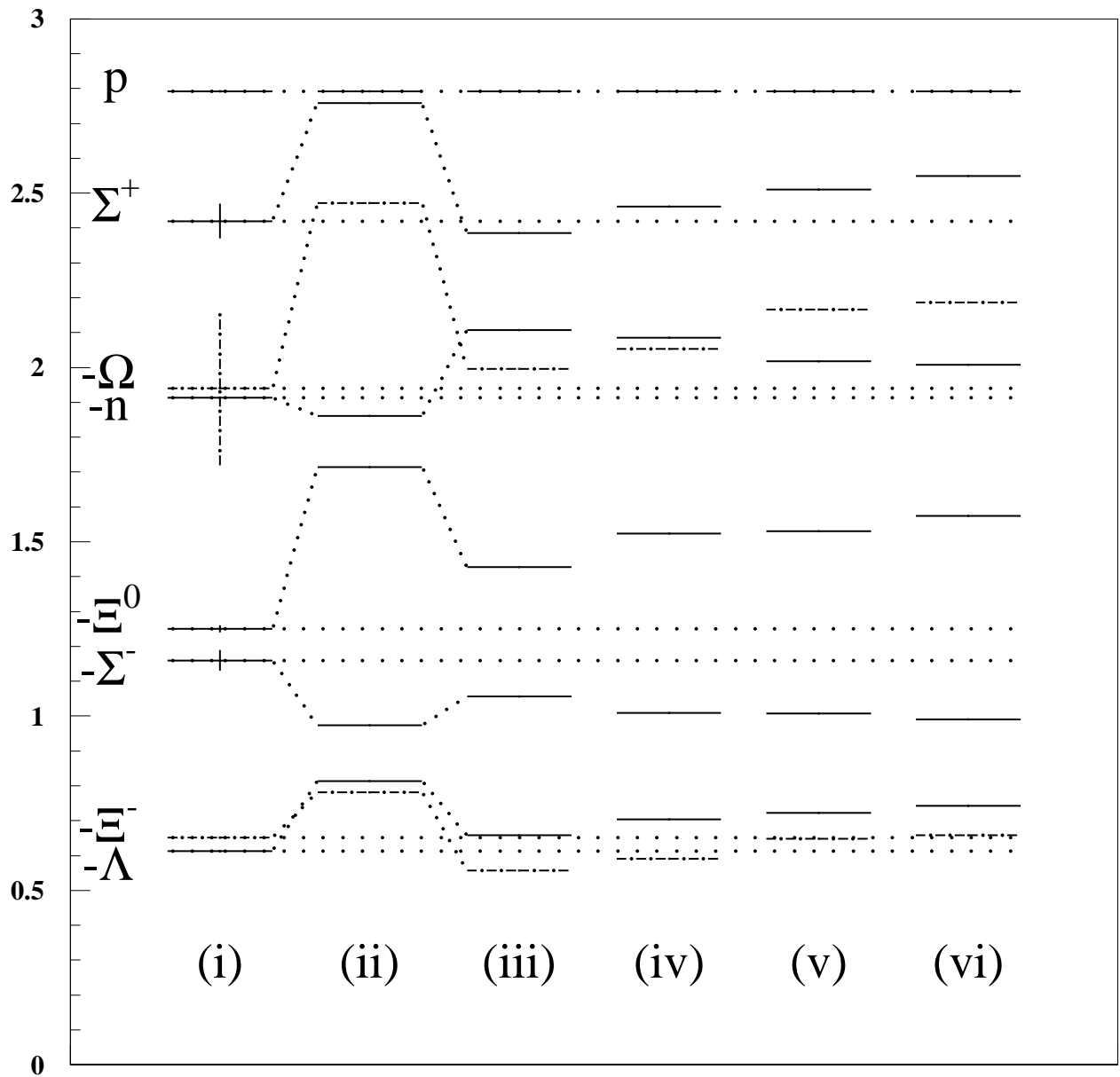
{\bf Fig.~3} \quad The effects of mesons and comparison with the cloudy bag.\\
(i) Experimental magnetic moments; \\ (ii) no mesons (CBM or CDM);\\
(iii) CDM, parameter set as Fig.~2(iii), pions only;\\
(iv) CDM with all mesons; \\(v) CBM, $r_{\rm bag}=1.0fm$, $m_s=144$, pions
only;
\\(vi) CBM, all mesons.\\

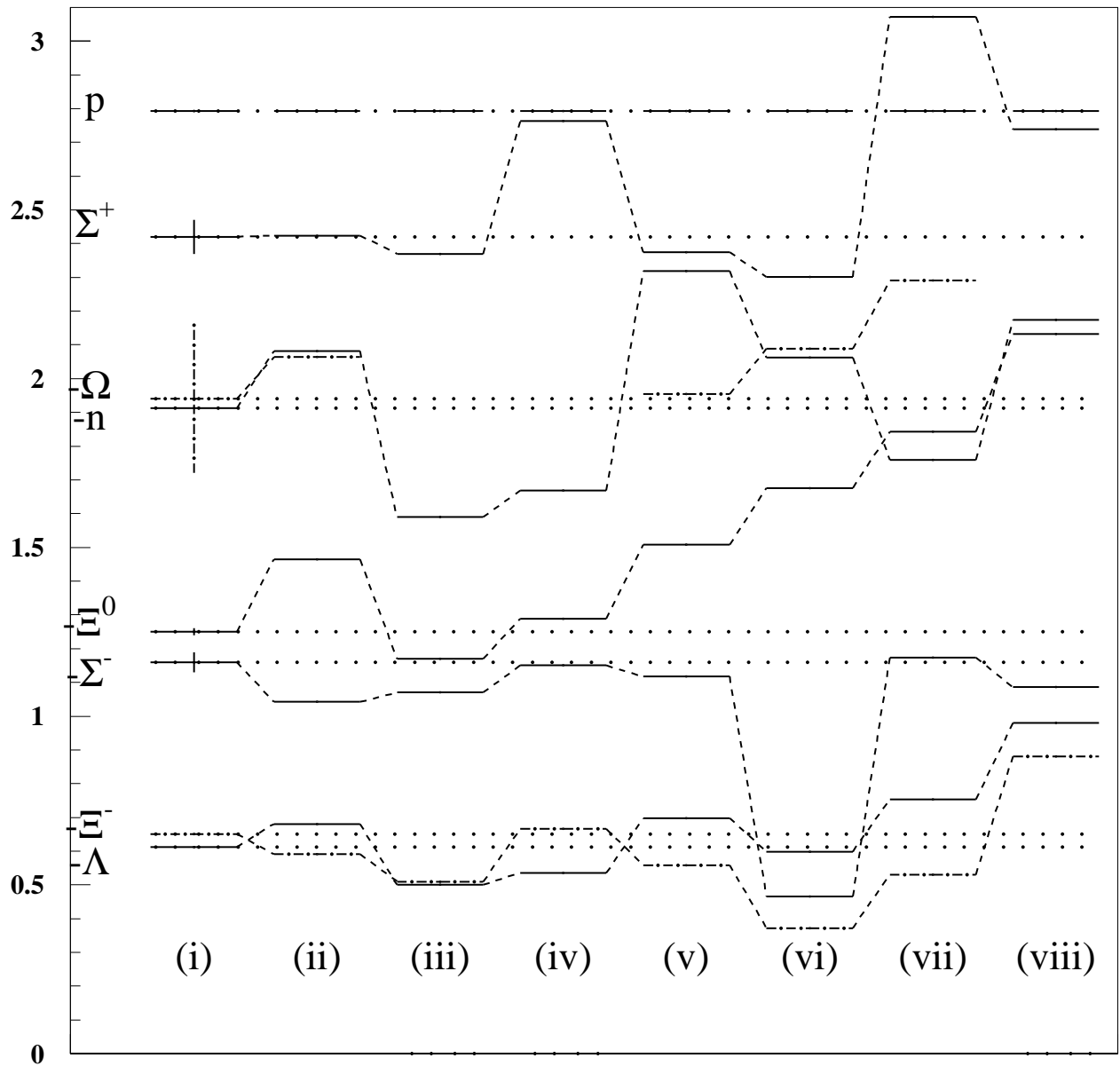
{\bf Fig.~4} \quad  Comparison with other models.  For more details, see
text.\\
(i) Experimental magnetic moments;\\ (ii) CDM, as Fig.~3(iii);\\
(iii) Lattice QCD\ \cite{lattice magmom};\\
(iv) Non-relativistic quark model with pion exchange currents\ \cite{Wagner};\\
(v) Skyrme model (4th order) in the slow rotor approximation\
\cite{Schwesinger};\\
(vi) Chiral bag model, $r_{\rm bag}=1.0$~fm\ \cite{Hong};\\
(vii) Skyrme model in the bound state approximation\ \cite{Oh};\\
(viii) Skyrme model (Yabu-Ando diagonalisation)\ \cite{Park}.
\end{document}